\documentclass[twocolumn,aps,floats,floatfix,prl,superscriptaddress,tightenlines]{revtex4}
\usepackage{graphicx}
\usepackage{bm}
\usepackage{epsfig}
\usepackage{pstricks}
\usepackage{amsmath}
\usepackage{hyperref}

\newcommand{\beq}{\begin{equation}}
\newcommand{\eeq}{\end{equation}}
\newcommand{\bea}{\begin{eqnarray}}
\newcommand{\eea}{\end{eqnarray}}
\newcommand{\OMIT}[1]{}

\begin{document}
\title{Smoking Guns for On-Shell New Physics at the LHC} 
\author{ Christian Arnesen 
}
\address{Carnegie Mellon University Dept. of Physics, Pittsburgh PA  15213, USA}
\author{ Ira Z. Rothstein
}
\address{Carnegie Mellon University Dept. of Physics, Pittsburgh PA  15213, USA}
\author{ Jure Zupan}
\address{Theory Division, Department of Physics, CERN, CH-1211 Geneva 23,
Switzerland}
\address{J.~Stefan Institute, Jamova 39, 1000 Ljubljana, Slovenia}
\address{Faculty of mathematics and physics, University of Ljubljana, Jadranska 19, 1000 Ljubljana, Slovenia}

\begin{abstract}

Using Tevatron bounds we derive upper limits on the LHC Higgs production rate under the assumption that no beyond the Standard Model (BSM) particles are being produced near their mass shell. A violation of these limits would constitute a smoking gun for light BSM particles. Furthermore, we demonstrate how \(R_T\), the ratio of the partially integrated Higgs transverse momentum distribution to the inclusive rate, can also be used as a probe of light BSM particles. This ratio is insensitive to heavy virtual effects and can be well-approximated by its SM value, \emph{i.e.} it is model independent. The perturbative expansion for \(R_T\) has reduced renormalization scale dependence, at the order of $5\%$ at next-to-leading order in QCD, due to a cancellation of Wilson coefficients. A deviation from the SM value implies that light BSM particles are being produced near their mass shell. We discuss a possible loophole to this conclusion, namely the existence of a non-perturbative, CP violating sector that couples to the Higgs. We use a toy model with colored scalars to demonstrate how the model independent prediction for $R_T$ is approached as the mass of the scalar becomes large.
\end{abstract}

\maketitle
The Higgs sector of the Standard Model (SM), responsible for electroweak symmetry breaking, has remained hidden from observation. At 95\% confidence level  LEP data constrain \(m_H>114.4\) GeV~\cite{Barate:2003sz}. Direct searches at the Tevatron have put upper limits on Higgs event rates at $1-7$ times the SM value for Higgs masses in the range $110-200$ GeV~\cite{Group:2008ds}. At the Large Hadron Collider (LHC) particles many times more massive than the top quark will be kinematically accessible. We remain hopeful that BSM particle production will manifest itself in resonances or shoulders of distributions, but we may not be that fortunate, particularly in the early stages of running. New-physics (NP) signals must be separated from large SM backgrounds in the complicated environment of a hadron collider, and NP discovery may remain elusive. Here we will explore a modest strategy in which we ask, ``How can we discern if deviations from the SM are due to  particles being produced on-shell?"  We approach this question through the framework of effective field theory (EFT).

The premise of EFT is that our theoretical description of low energy observables need not include heavier particles as dynamical degrees of freedom. Instead we can approximate virtual exchange of the massive particles as a set of local contact interactions. The approximation is a power series expansion in \(p^2/\Lambda^2\) where $p^2$ is a typical kinematic invariant in the process and the EFT ``cutoff'' \(\Lambda\) is the mass scale of the exchange. For instance, if the interaction is mediated by pair-produced particles of mass $m$ then \(\Lambda\sim 2 m\). At each order in the expansion, the set of local operators is the most general one consistent with low energy symmetries, whatever the massive dynamics may be. Absent additional model assumptions, the Wilson coefficients of the local operators must be determined phenomenologically. Then the EFT predicts concrete relationships between collider observables such as production rates or branching ratios in related channels.

In this letter, we make model independent EFT predictions for new physics in Higgs production through gluon fusion, which starts at one loop in the SM and will be the dominant Higgs production mechanism at the LHC. The generality of the EFT approach ensures that any massive BSM extension can be accommodated. Conversely, deviations from the EFT predictions signify {\it model independently} that light particles with \(m\sim p\) are being produced. We will demonstrate this for the case of Higgs production, where the presence of new particles with masses $m^2\sim m_H^2+p_T^2$ can be probed.

A complete basis of dimension-six EFT operators for the Standard Model was first constructed in~\cite{Buchmuller:1982ye}. At leading order in the power expansion, there is just one CP even and one CP odd operator that can modify the Higgs-glue interaction 
\beq\label{Leff}
{\cal L}_{\rm eff}=C_G H G_{\mu \nu}^a G^{a\mu \nu}+\tilde C_G H \tilde G_{\mu \nu}^a G^{a\mu \nu}, 
\eeq
where \(G_{\mu\nu}^a\) is the gluon field strength and \(\tilde G_{\mu\nu} = \epsilon_{\mu\nu\rho\sigma}G^{\rho\sigma}/2\). For $m_H\lesssim 200~$GeV the SM Higgs-glue interaction is point-like to an excellent approximation due to the large top-quark mass and can be well-described by the same effective Lagrangian, Eq.~(\ref{Leff}). The infinite top mass approximation has been used extensively in the literature since it greatly simplifies the calculation of radiative corrections~\cite{Anastasiou:2005qj,Glosser:2002gm,Ravindran:2003um}. In this limit, a top loop contributes $C_G^{\rm SM}=\alpha_S/(12 \pi v)$ to the effective Higgs-gluon coupling at leading order. Heavy NP will modify the value of $C_G=C_G^{\text{SM}}+C_G^{\text{NP}}$ and, if CP violating, generate $\tilde C_G$. If we assume that the new physics does not get mass from electroweak symmetry breaking, then the magnitude of $\tilde C_G$ is bounded by upper limits on the electric dipole moment of the neutron $d_n<2.9\times 10^{-26}$e-cm~\cite{edm}. For the moment we will neglect CP violating effects and return to this possibility in the conclusions.

\begin{figure}[!t]
\centerline{\scalebox{0.75}{\includegraphics{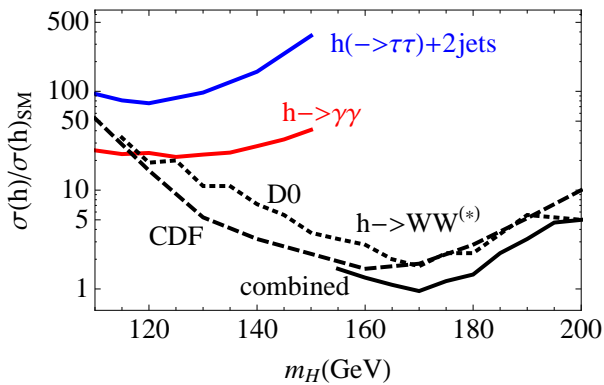}}}
\vskip-0.3cm
\caption[1]{$95\%$ CL upper bound on gluon-fusion Higgs production at the LHC relative to the SM value, $\sigma(H)/\sigma(H)_{\rm SM}=|C_G/C_G^{\rm SM}|^2$, coming from Tevatron searches at $2.0-3.0~{\rm fb}^{-1}$:  $H(\to \gamma\gamma)$ (red line)~\cite{D0gammagamma}; $H(\to \tau\tau)$+2 jets (blue line)~\cite{CDFtautau}; $H\to W W^{(*)}$ (dotted line)~\cite{D0WW,CDFWW}; combined analysis (solid line)~\cite{Bernardi:2008ee}.}
\label{tev} 
\vskip-0.5cm
\end{figure}

\begin{figure}[!t]
\centerline{\scalebox{0.90}{\includegraphics{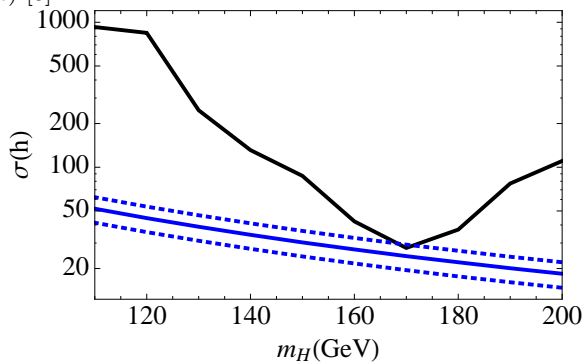}}}
\vskip-0.3cm
\caption[1]{95\% CL upper bounds on the Higgs production cross section at the LHC from Tevatron Higgs searches~\cite{D0gammagamma,CDFtautau,D0WW,CDFWW,Bernardi:2008ee}. The upper bound includes the error in the theoretical cross section at NNLO, which we conservatively approximate to be $20\%$. The lower lines are the SM prediction with associated errors.}
\label{LHCpred} 
\vskip-0.5cm
\end{figure}

Searches at the Fermilab CDF and D0 experiments put upper bounds on Higgs production rates~\cite{Group:2008ds,Bernardi:2008ee}, which we translate into bounds on $C_G$. The relevant production mechanisms are $p\bar p\to H(\to \gamma\gamma)$~\cite{D0gammagamma}, $p\bar p\to H\to W W^{(*)}$~\cite{D0WW,CDFWW} and $p\bar p\to H(\to \tau\tau)$+2 jets~\cite{CDFtautau}. The first two channels are gluon fusion dominated. Note that Higgs production with an associated weak boson, which dominates the combined Tevatron Higgs bounds for small Higgs masses~\cite{Group:2008ds}, are not sensitive to glue-Higgs interactions. The resulting $95\%$ confidence level (CL) upper limits on $|C_G/C_G^{\rm SM}|^2$ are shown in Fig.~\ref{tev}, assuming SM decays for the Higgs boson.  In Fig.~\ref{LHCpred} we have translated Fig.~\ref{tev} into bounds on $\sigma(pp\to H +X)$ at the LHC at next-to-next-to-leading order (NNLO) in the expansion in $\alpha_s$ utilizing  {\tt Fehip}~\cite{Anastasiou:2005qj}. A measurement exceeding these EFT bounds is a smoking gun for light BSM particles. 

\begin{figure}[!t]
\centerline{\scalebox{0.75}{\includegraphics{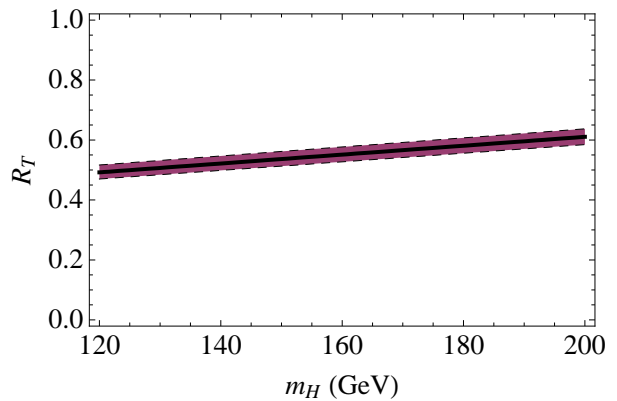}}}
\vskip-0.3cm
\caption[1]{ Model independent value of $R_T$ from Eq.~(\ref{R}) at NLO in \(\alpha_S\) as a function of Higgs mass. Vector boson fusion contributions are excluded. The uncertainty band is approximated by varying the renormalization scale by a factor of two.
}
\label{R-fig} 
\vskip-0.5cm
\end{figure}

There are other model independent signatures for BSM particle production that do not rely on Tevatron input and can thus be improved at LHC with increasing statistics. The observable that we will focus on is the ratio of Higgs production cross section at large transverse momenta to the totally inclusive rate,
\beq\label{R}
R_T\equiv\frac{\sigma(H \, : \, p_T^H>p_T^{\rm min})}{\sigma(H)}\,,
\eeq
with a lab-frame rapidity cut \(|y_H|<y^{\rm max}_H\) in both numerator and denominator. The transverse Higgs production rate is interesting in its own right as a Higgs search channel~\cite{Ellis:1987xu,Abdullin:1998er}. The ratio Eq.~(\ref{R}) has several favorable properties: (i) independence from  Higgs branching ratios, (ii) reduced perturbative uncertainty relative to the individual cross sections, and most importantly (iii) heavy NP cannot change its value from the SM one. These are {\it model independent} predictions of the EFT. Figure~\ref{R-fig} shows the EFT prediction of this ratio calculated at NLO in \(\alpha_S\) using {\tt Fehip}~\cite{Anastasiou:2005qj} for a range of Higgs masses in the large-$m_t$ limit. To have a clear EFT model independent interpretation of $R_T$, the vector boson fusion contributions should be subtracted from the measurement. This can be done precisely at NLO using {\tt VBFNLO} \cite{Hankele:2006ma}.

The error band in Fig.~\ref{R-fig} reflects the perturbative uncertainty, estimated from varying the renormalization scale by a factor of two in each direction. While the variation in the transverse and inclusive rates individually are of order $20 \%$, the ratio is relatively insensitive with only $ 5 \%$ variation. A simple way to understand this reduction in errors can be seen from an EFT point of view.  There are two sources of $\alpha_S$ in the calculation: the matrix element and the matching coefficient. In the large-top-mass limit all of the $\alpha_S$ dependence in the ratio stemming from the matching coefficient cancels.  Errors due to parton distribution function (PDF) uncertainties also cancel to large extent in the ratio. A calculation of the cross sections using several different PDFs reveals that these errors are negligible compared to the perturbative errors~\cite{Ravindran:2003um}. When $|C_G^{\rm NP}| \lesssim |C_G^{\rm SM}|$, one needs to consider other corrections to the large-top-mass SM prediction beyond perturbative ones. Kinematic power corrections due to finite top-quark mass are small, \(\lesssim\)10\%, for \(m_h,\,p_T^{\min}\lesssim\)~200~GeV~\cite{Glosser:2002gm}, as are contributions with a $b$ quark in the higgs-glue-glue loop. The two-loop electroweak corrections modify the $gg\to h$ cross section at the $5\%-8\%$ level~\cite{Degrassi:2004mx}, but are likely smaller for the $R_T$ ratio.
 
What could we conclude from a deviation from Fig.~\ref{R-fig}? One strong possibility would be that such a deviation would arise as a consequence of light colored particles. We might hope to see these particles in jet production cross sections, but absent knowledge of their widths or decay products this may be difficult. However, we could learn more about their masses by altering the cuts on the Higgs $p_T$ distribution and seeing how $R_T$ changes. As the cut increases we would expect the deviations to grow. With sufficient statistics we should be able to find the value, $p_T$, at which the distribution begins to deviate from the SM and infer the presence of on-shell particles with mass $\sim \sqrt{m_H^2+p_{T}^{2}}$. Note that in principle one could also get deviations without light BSM particles if some other higher dimension operators became relevant, for example, four-quark operators. PDF suppression is large though for quark initiated Higgs production, and LEP bounds constrain the Wilson coefficients of these operators. So this possibility is excluded.

\begin{figure*}[!t]
\includegraphics[width=7.2cm]{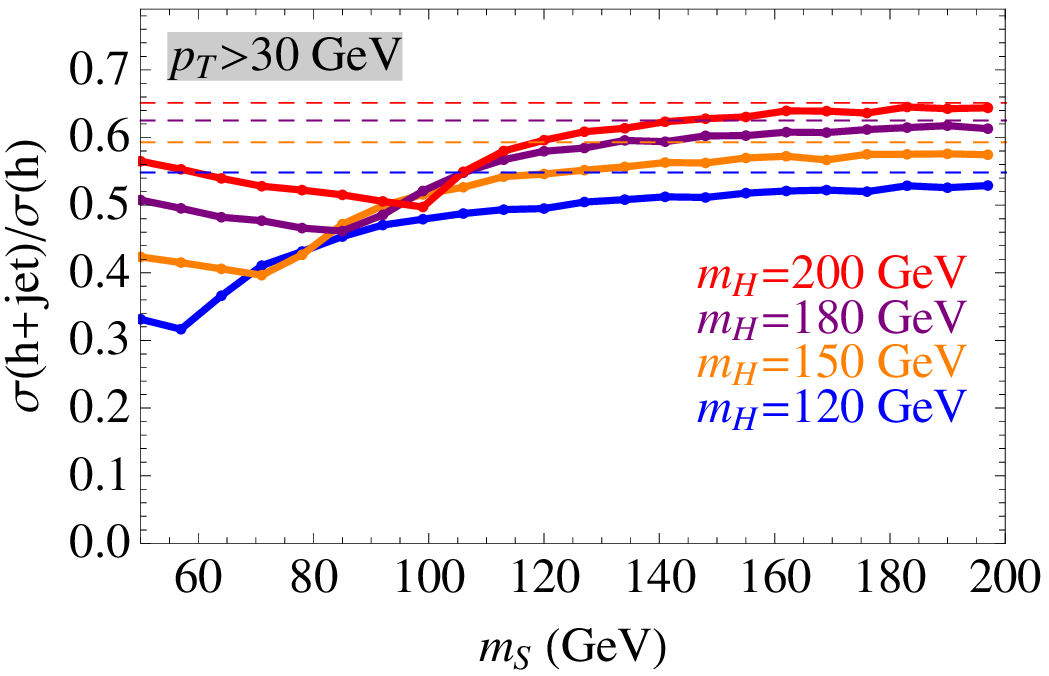}~~~~~~~~~~~~\includegraphics[width=7.2cm]{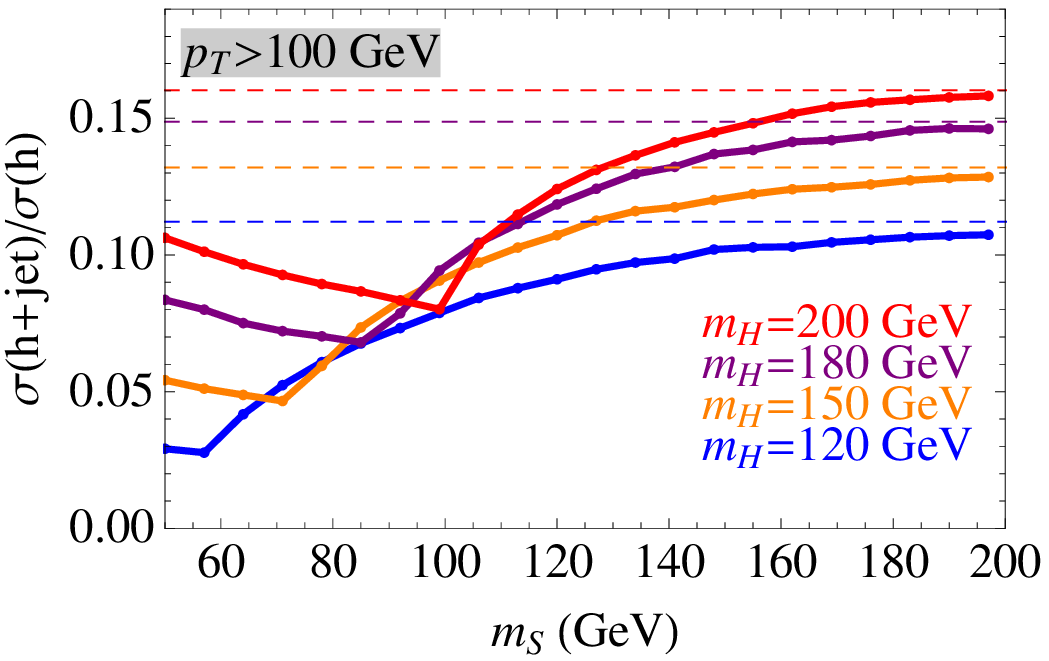}
\vskip-0.3cm
\caption[1]{$m_S$ dependence of $R_T$ for $p_T^H>30$~GeV (left) and $p_T^H>100$~GeV (right) for the toy model (solid lines) and for the SM (dashed lines) at LO in \(\alpha_S\).
The values of Higgs mass are as indicated. Calculations aided by {\tt FeynArts}, {\tt FormCalc} and {\tt LoopTools} packages~\cite{Hahn:2000kx}. No $y_H$ cut is imposed and $\kappa$ has been set to unity.}
\label{plotRscalar} 
\vskip-0.5cm
\end{figure*}

{\sl Toy Model}. Let us give a concrete example of how $R_T$ can deviate from the EFT prediction when there are new light particles in the theory. As a toy-model example we extend the SM by adding a color-octet weak-singlet scalar \(S\) that couples to the Higgs field through a trilinear term
\beq {\cal L}=
-\frac{\kappa}{2} v H S^a S^a+\cdots.
\eeq
after electroweak symmetry breaking, where $v\simeq246$~GeV is the Higgs vacuum expectation value and ``\(+\cdots\)" refers to other terms in the scalar potential that are not relevant to our analysis. The colored scalars contribute virtually through co-annihilation, $pp\to 2S\to H$, as well as through on-shell production $pp\to H+2S$. For the regions of parameter space explored here, the latter is numerically subleading due to gluon PDF suppression at large momentum fractions. The value of $R_T$ in the SM and in the toy-model extension are shown in Fig.~\ref{plotRscalar} for Higgs momentum cuts $p_T^{\rm min}=30$~GeV and $p_T^{\rm min}=100$~GeV and a range of octet-scalar masses.

If the scalar's mass is larger than the typical partonic center-of-mass energy (set by the Higgs mass and momentum), then its effect is simply to shift the Wilson coefficient $C_G$ in Eq.~\eqref{Leff}. At leading order in $\alpha_S$,
\beq
C_G^{NP}=\frac{\kappa \alpha_s N_cv}{96 \pi m_S^2}. 
\eeq
In this scenario, the transverse and inclusive production rates individually differ from their SM values by a factor \(|(C_G^{\rm SM}+C_G^{\rm NP})/C_G^{\rm SM}|^2\), but their ratio is unchanged.  Qualitatively we can expect that deviations of $R_{T}$ from the EFT prediction will only arise when the scalar is sufficiently light. This is seen in Fig.~\ref{plotRscalar} where $R_{T}$ approaches the SM value when \(m_S\) is large. Furthermore as the cut on the Higgs $p_T$ is increased we would expect this deviation to grow since a larger partonic center-of-mass energy leads to enhanced power corrections. This is a generic feature of all NP models that could modify $R_T$. Note that since the Higgs branching ratios cancel in the ratio, modification can only come from the Higgs production. In more realistic SM extensions, such as supersymmetry or models with extra singlet scalars, \(R_T\) will generally differ from its SM value with the magnitude of the deviation depending on the specific model parameters.

A benefit of the $R_T$ ratio is that it does not rely on the decay properties of the new resonances. In our toy model the octet scalars would also be pair produced (without a Higgs), modifying the two-jet production cross section and other QCD observables. Searches in these channels however would depend on the scalar decay products, and dedicated studies may be needed~\cite{Dobrescu:2007yp}. It could be that separating these signals from background is prohibitively challenging, for example if there is some (not necessarily fine) degeneracy with another final state as in light stop scenarios of the MSSM \cite{Carena:2008rt}.

Violations of observable predictions of the EFT may be searched for in other channels dominated by gluon fusion as well, for instance $h$+2 jets. With increasing statistics, the EFT approach outlined here can also be used in channels that proceed at tree level in the SM, in which case EFT contributions would be a subleading effect at low energies. With high $p_T$ cuts in TeV range, these effects may become leading, offering another way to search model independently for TeV resonances. We leave these possibilities for future studies. 

{\sl Conclusions}. We showed that even if the Tevatron does not discover the Higgs boson, the tightening of constraints on its production may facilitate the search for on-shell BSM particles at the LHC. The Tevatron and LHC results are complementary because the collisions occur at different energies. For example, if the EFT description applies only to Tevatron results, the LHC bounds in Fig.~\ref{LHCpred} may be violated, acting as a smoking gun for on-shell particle production. Other observables can be used in the indirect searches for on-shell new particles. Figure~\ref{R-fig} gives a model independent prediction for the ratio of the Higgs production rate at large transverse momenta to the inclusive rate. Should the data disagree significantly with this prediction then one may conclude that new particles have been produced in the collisions and presumably escaped detection either due to backgrounds or the nature of the decay products. This reasoning assumes that the CP violating Wilson coefficient $\tilde C_G$ in Eq.~\eqref{Leff} is negligible. This is true if the off-shell new physics is perturbative, as $\tilde C_G$ can arise only at two loops and is thus much smaller than $C_G^{\rm SM}$. Furthermore, the presence of $\tilde C_G$ can be bounded model independently from experiment by measuring $R_T$ at several Higgs $p_T$ cuts. Namely, $R_T\propto(C_G^2+\tilde C_G^2)/(C_G^2+2 \tilde C_G^2)$, where the undisplayed known kinematical factor depends on the Higgs $p_T^{\rm min}$. CP violating measurements can also be made in the two jet sector \cite{zep}.

In the case of smaller deviations (say less than 50\%), a more refined analysis would be necessary to determine if the cause is light new physics. Electroweak corrections would have to be included along with the $b$-quark loop effects. When these effects become important  $R_T$ is no longer model independent. Nonetheless one could still use the EFT strategy discussed here to search for light new physics, by first extracting $C_G$ from the inclusive rate and then using it to predict the cut rate.

We have presented a model with a color octet scalar which shows how the prediction in Fig.~\ref{R-fig} is approached as the scalar mass gets large. It is important to note, however, should the data fail to produce the correct value of $R_T$, it does not necessarily imply that the new particles being produced have non-trivial color quantum numbers. Nonetheless, a natural place to look for these new particles would be in jet production rates, although detecting such effects will clearly depend upon the nature of the decay process of the new particles.

Finally, we would like to emphasize that even if new BSM particles are discovered in channels not involving the Higgs, $R_T$ remains an interesting observable in its own right. For one, it is under better theoretical control due to the cancellation of QCD corrections mentioned in the body of the paper. In addition,  the discovery of new particles not involved in Higgs production, does not imply that $R_T$ will deviate from its SM value.  In fact if $R_T$ did not deviate from  its SM value in this scenario, then this would provide very useful information in discerning models.

{\sl Acknowledgements.} The authors would like to thank T. Ferguson, B. McElarth, T. Hahn, S. Jaeger, T. Rejec, J. Russ, J. Wells and K. Yorita. 

The work of C.A. and I.Z.R. is supported by \uppercase{DOE} contracts \uppercase{DOE-ER}-40682-143 and \uppercase{DEAC02-6CH03000}.

\end{document}